\documentclass[superscriptaddress,showkeys,prd]{revtex4}
\usepackage{amsmath}
\usepackage{amssymb}
\usepackage{graphicx}
\usepackage{color}
\usepackage{ulem}

\begin{document}

\title{
Quantum monopole via Heisenberg quantization}

\author{
Vladimir Dzhunushaliev
}
\email{v.dzhunushaliev@gmail.com}
\affiliation{
IETP, Al-Farabi KazNU, Almaty, 050040, Kazakhstan
}
\affiliation{
Dept. Theor. and Nucl. Phys., KazNU, Almaty, 050040, Kazakhstan
}
\affiliation{
Institute of Physicotechnical Problems and Material Science of the NAS
of the Kyrgyz Republic, 265 a, Chui Street, Bishkek 720071,  Kyrgyz Republic
}
\affiliation{
Institute of Systems Science, Durban University of Technology, P. O. Box 1334, Durban 4000, South Africa
}

\begin{abstract}
Using a non-perturbative quantization method originally due to Heisenberg we obtain {\it quantum} monopole solutions and {\it quantum} flux tube solutions for the SU(3) strong interaction gauge theory. For the quantum monopole solution we find that the radial chromomagnetic field decreases exponentially with a scale set by the effective gluon mass. The quantum flux tube solution stretches between a monopole and anti-monopole and has a longitudinal chromomagnetic field. Both solutions exhibit characteristics of the Meissner effect and are conjectured to have a connection to the confinement phenomenon. 
\end{abstract}


\maketitle

\section{Introduction}

At the present time the most interesting problems in quantum chromodynamics, such as confinement, calculation the hadrons masses, and many others, can not be solved, since for their solution it is necessary to have nonperturbative calculation methods in quantum field theory. One of the methods of overcoming this problem is numerical calculations on a lattice. One of the results of these calculations was the establishment of the fact that monopoles play an important role in quantum chromodynamics. The condensation of monopoles leads to compression of electric fields in a tube stretched between sources of colored fields. The monopoles in quantum medium can be seen after fixing the Abelian calibration. Unfortunately, the microscopic structure of these monopoles remains unclear and apparently can only be investigated in nonperturbative analytic calculations. In this paper we obtain a monopole and flux tube solutions in the approximation of two equations of nonperturbative quantization a la Heisenberg. This allows us to study in detail the microscopic structure of such a quantum monopole which is distinct from the 't Hooft - Polyakov monopole.

Classical non-Abelian gauge theories have a rich family of classical solutions. One of the best known of these solutions are the 't Hooft-Polykov monopole solutions \cite{thooft} \cite{polyakov}. The original version of 't Hooft-Polykov monopole solution was found in the context of an SU(2) gauge theory coupled to a scalar field have a $\lambda \phi ^4$ symmetry breaking potential. The Lagrange density for this system was
\begin{equation}
	\label{lagrange-thp}
	{\cal L} = - \frac{1}{4} F^a _{\mu \nu} F^{a \mu \nu} + \frac{1}{2} (D^\mu \phi ^a) (D_\mu \phi ^a) - \frac{\lambda}{4} (\phi ^a \phi ^a - v^2) ~,
\end{equation}   
where   $F^a_{\mu \nu} = \partial_\mu A^a_\nu - \partial_\nu A^a_\mu + g \epsilon^{abc} A^b_\mu A^c_\nu$ is the SU(2) field strength tensor; $A^a_\mu$ is the gauge potential; $a, b, c = 1, \ldots , 3$ are the SU(2) indices; $g$ is the coupling constant; $\epsilon^{abc}$ are the structure constants for the SU(2) gauge group; $D_\mu \phi ^a = \partial _\mu \phi ^a - g \epsilon ^{abc} A^b _\mu \phi ^c$ is the SU(2) covariant derivative; the last term proportional to $\lambda$, is a Higgs symmetry breaking potential which gives the scalar field a vacuum expectation value of $v = \sqrt{\langle \phi ^a \phi ^a \rangle}$. The classical equations of motion resulting from \eqref{lagrange-thp} are
 \begin{eqnarray}
	D_\nu {F}^{a  \mu\nu} &=&  - g \epsilon ^{abc} \phi ^b (D _\mu \phi ^c ) 
\label{Fuv} \\
	(D^  \mu D_\mu \phi ^a ) &=& - \lambda \phi ^a (\phi ^b \phi ^b - v^2 ).
\label{phi-2}
\end{eqnarray}  
By making the radial ansatz for the scalar and gauge fields of the form
\begin{equation}
	\phi ^a = \frac{r^a}{g r^2} H(r) ~~~~;~~~~ A^{i a} = - \epsilon ^{a i j} \frac{r^j}{g r^2}[1 - K(r)] ~,
\end{equation}
the equations \eqref{Fuv} \eqref{phi-2} take the form
\begin{eqnarray}
	x^2 \frac{d^2 K}{dx^2} &=& K H^2 + K(K^2 -1) \label{K-eqn} ~, \\
	x^2 \frac{d^2 H}{dx^2} &=& 2 K^2 H +\frac{\lambda}{g^2} H(H^2 - x^2) \label{H-eqn} ~.
\end{eqnarray}
where $x=v g r$ is the re-scaled radial coordinate. These equations \eqref{K-eqn} \eqref{H-eqn} where shown to have classical solutions which at large distances behaved as Dirac monopoles and at short distances had smooth, non-singular fields which exponentially merged with the large distance monopole form \cite{thooft} \cite{polyakov}. Furthermore these  monopole field configurations were stable due to their non-trivial topological structure.  

While the classical 't Hooft-Polyakov solutions have generated a lot of interest and work, it is fair to say that they have not yet made an experimentally verified contribution to particle physics. From the theoretical side this can be attributed to two points: (i) they are classical solutions which do not take quantization into account and (ii) they require scalar fields of a type that have not yet been observed (i.e. scalar fields which couple the SU(3) gauge fields). In this work we will apply a quantization method due to Heisenberg \cite{heis} to the SU(3) non-Abelian gauge theory of the strong interaction. 

The paper is organized as follows: in Section \ref{sec2} we consider two equations approximation, in Section \ref{sec3} a quantum monopole is obtained, in Section \ref{sec4} an infinite flux tube is obtained, in Section \ref{sec5} Meissner effect is discussed. 

\section{Two equations approximation method for non - perturbative quantization in QCD}
\label{sec2}

The method of non -perturbative quantization \`{a} la Heisenberg is that we have to write operator Yang - Mills equations \cite{Dzhunushaliev:2016svj} 
\begin{equation}\label{1-10}
	D_\nu \widehat {F}^{a \mu\nu} = 0,
\end{equation}
where
$\hat F^B_{\mu \nu} = \partial_\mu \hat A^B_\nu - \partial_\nu \hat A^B_\mu +
g f^{BCD} \hat A^C_\mu \hat A^D_\nu$ is the field strength operator;
$\hat A^B_\mu$ is the gauge potential operator; $B, C, D = 1, \ldots , 8$ are the SU(3) color indices; $g$ is the coupling constant; $f^{BCD}$ are the structure constants for the SU(3) gauge group. 

This equations set is equivalent to an infinite set of equations  for all Green functions
\begin{eqnarray}
	\left\langle
		D_\nu \widehat {F}^{A \mu\nu} (x)
	\right\rangle &=& 0 ,
\label{1-20}\\
	\left\langle
		\hat A^{B_1}_{\alpha_1} (x_1)
		D_\nu \widehat {F}^{A \mu\nu} (x)
	\right\rangle &=& 0 ,
\label{1-30}\\
	\left\langle
		\hat A^{B_1}_{\alpha_1} (x_1) \hat A^{B_2}_{\alpha_2} (x_2)
		D_\nu \widehat {F}^{A \mu\nu} (x)
	\right\rangle &=& 0 ,
\label{1-40}\\
	\ldots &=& 0	,
\label{1-50}\\
	\left\langle
		\hat A^{B_1}_{\alpha_1} (x_1) \ldots \hat A^{B_n}_{\alpha_n} (x_n)
		D_\nu \widehat {F}^{A \mu\nu} (x)
	\right\rangle &=& 0
\label{1-60}\\
	\ldots &=& 0 .
\label{1-70}
\end{eqnarray}
The set of equations \eqref{1-20} -- \eqref{1-70} is known as Dyson -- Schwinger equations. But usually these equations are written as Feynman diagrams. We will not use this notation to emphasize the nonperturbative nature of the calculations carried out in this article. 

Let us compare our situation with infinite equations set \eqref{1-20} -- \eqref{1-70} with turbulence modeling. In reference \cite{Wilcox} (sections 2.3 and 2.4) it is shown that if we make some realistic physical assumptions about high cumulants with velocity and pressure then similar infinite equations set can be reduced to two equations: the averaged Navier -- Stokes and Reynolds-Stress equations. In turbulence modeling the problem of cutting off infinite equations set 
to such two equations is known as the closure problem. 

In a similar way in Ref. \cite{Dzhunushaliev:2016svj} it is shown that infinite equations set \eqref{1-20} -- \eqref{1-70} can be approximately reduced to two equations describing 2- and 4- point Green functions, namely, 
\begin{eqnarray}
	\tilde D_\nu F^{a \mu \nu} - \left[
		\left( m^2 \right)^{ab \mu \nu} -
		\left( \mu^2 \right)^{ab \mu \nu}
	\right] A^b_\nu &=& 0 ,
\label{1-80}\\
	\Box \phi - \left( m^2_\phi \right)^{ab \mu \nu} A^a_\nu A^b_\mu \phi -
	\lambda \phi \left( M^2 - \phi^2  \right) &=& 0,
\label{1-90}
\end{eqnarray}
where $\tilde D_\mu$ is the gauge derivative of the subgroup $SU(2)$; 
$\left( m^2 \right)^{ab \mu\nu} \propto \phi^2$,
$\left( \mu^2 \right)^{ab \mu \nu}$  and $\left( m^2_\phi \right)^{ab \mu \nu}$ are quantum corrections coming from the dispersions of the operators 
$\widehat{\delta A}^{a \mu}$ and $\hat A^{m \mu}$: 
\begin{eqnarray}
	\left( m^2 \right)^{ab \mu\nu} &=& - g^2 \left[
		f^{abc} f^{cpq} G^{pq \mu\nu} -
		f^{amn} f^{bnp} \left(
			\eta^{\mu \nu} G^{mp \phantom{\alpha} \alpha}_{\phantom{mn} \alpha} -
			G^{mp \nu \mu}
		\right)
	\right] ,
\label{1-100} \\
	\left( \mu^2 \right)^{ab \mu \nu} &=& - g^2 \left(
		f^{abc} f^{cde} G^{de \mu \nu} +
		\eta^{\mu \nu} f^{adc} f^{cbe} G^{de \phantom{\alpha} \alpha}_{\phantom{de} \alpha} +
		f^{aec} f^{cdb} G^{ed \nu \mu}
	\right) ,
\label{1-110} \\
	\left( m^2_\phi \right)^{ab \mu \nu} &=&
	g^2 f^{amn} f^{bnp} \frac{
		G^{mp \mu \nu} - \eta^{\mu \nu} G^{mp \alpha}_{\phantom{mp \alpha} \alpha}
	}{G^{mm \alpha}_{\phantom{mm\alpha} \alpha}} .
\label{1-120}
\end{eqnarray}
$a, b,c = 2,5,7$ are the $SU(2)$ color indices; and $m,n = 1,3,4,6,8$ are coset indices. Eq.(\ref{1-80}) describes the 
$\left\langle \hat A^{a \mu} \right\rangle \in SU(2)$ degrees of freedom that have non-zero expectation values, and Eq.\eqref{1-90} describes the coset $\hat A^m_\mu \in SU(3) / SU(2)$ degrees of freedom with zero expectation values, i. e.,
\begin{eqnarray}
	\hat A^{a \mu} &=& \left\langle \hat A^{a \mu} \right\rangle +
	i \widehat{\delta A}^{a \mu} ,
\label{1-130}\\
	\left\langle \hat A^{m \mu} \right\rangle &=& 0 .
\label{1-140}
\end{eqnarray}
2-point Green functions for the gauge fields $\widehat{\delta A}^a_\mu \in SU(2) \times U(1)$ and for the coset  $\hat A^m_\mu \in SU(3) / SU(2)$ are defined as
\begin{eqnarray}
	G^{mn \mu \nu}(y,x) &=& \left\langle
		\hat A^{m \mu}(y) \hat A^{n \nu}(x)
	\right\rangle ,
\label{1-150}\\
	G^{ab \mu \nu}(y,x) &=& \left\langle
		\widehat{\delta A}^{a \mu}(y) \widehat{\delta A}^{b \nu}(x)
	\right\rangle .
\label{1-160}
\end{eqnarray}
We will use following approximation 
\begin{eqnarray}
	G^{mn \mu \nu}(y,x) &\approx& 
	\Delta^{mn} \mathcal A^\mu \mathcal A^\nu \phi(y) \phi(x) ,
\label{1-170}\\
	G^{ab \mu \nu}(y,x) &\approx& 
	\Delta^{ab}	\mathcal B^\mu \mathcal B^\nu .
\label{1-180}
\end{eqnarray}
where $\Delta^{ab} (a,b=2,5,7), \Delta^{mn} (m,n = 1,3,4,6,8)$ are constants; $\mathcal A_\mu \mathcal A^\nu, \mathcal B_\mu \mathcal B^\nu = \text{const}$. The condensate $\phi$ describes the averaged dispersion of quantum fluctuations of $\hat A^{m \mu}$ fields. 

In order to obtain equation \eqref{1-90} we assumed the following approximation for the 4--point Green function 
$
	G^{mnpq}_{\phantom{mnpq}\mu \nu \rho \sigma}(x, y, z, u) = 
	\left\langle
		\hat A^m_\mu(x) \hat A^n_\nu(y) \hat A^p_\rho(z) \hat A^q_\sigma(u)
	\right\rangle
$
\begin{eqnarray}
    G^{(4)} \approx \frac{\lambda}{4} \left(
        G^{(2)} - M^2  \
    \right)^2 .
\label{1-190}
\end{eqnarray}
Let us note that in our opinion the two equations approximation can be applied to the static case only. 

\section{Heisenberg's quantum monopole}
\label{sec3}

Our goal is to obtain a monopole-like solutions for equations \eqref{1-80} - \eqref{1-90}. The physical meaning of such a solution is as follows: the magnetic field is pushed out by the condensate $\phi$ and forms a monopole with an exponentially falling magnetic field. This is the Meissner effect in QCD. 

The ans\"atz for the SU(2) gauge fields is taken in the standard monopole form 
\begin{eqnarray}
	A^a_\mu &=& \frac{2}{g} \left[ 1 - f(r) \right] 
	\begin{pmatrix}
		0	& 0 & 0 & - \sin^2 \theta \\
		0	& 0 & \cos \phi & - \sin \theta \cos \theta \sin \phi \\
		0	& 0 & \sin \phi & \phantom{-} \sin \theta \cos \theta \cos \phi \\
	\end{pmatrix} , 
\label{2-10} \\
	\phi &=& \phi(r) 
\end{eqnarray}
here $a=2,5,7$ and $\mu = t,r,\theta,\phi$. 

We choose following matrix $\Delta^{AB}, (A,B = 1,2 , \ldots , 8)$ 
\begin{equation}
	\Delta^{AB} = \text{diag} \left( 
		\Delta_{11}, \delta_{2}, \Delta_{33}, \Delta_{44}, 
		 \delta_{5}, \Delta_{66}, \delta_{7}, \Delta_{88}
	\right) 
\label{2-20}
\end{equation}
with $\Delta_{66} = \Delta_{44}$, 
$\Delta_{44} + \Delta_{88} = \Delta_{11} + \Delta_{33}$ and vectors $\mathcal A_\mu, \mathcal B_\mu$ 
\begin{eqnarray}
	\mathcal A_\mu &=& \left( 
		\mathcal A_0, \mathcal A_1, 0, 0 
	\right) , 
\label{2-30}\\
	\mathcal B_\mu &=& \left( 
		\mathcal B_0, 0, 0, 0 
	\right)  .
\label{2-40}
\end{eqnarray}
Using $\Delta^{AB}$, $\mathcal A_\mu$ and $\mathcal B_\mu$ from \eqref{2-20} - \eqref{2-40} we obtain 
\begin{eqnarray}
	\left( m^2 \right)^{4b 3 \nu} A^b_\nu &=& 
	- \left( m^2 \right)^{5b 2 \nu} A^b_\nu = 
	\left( m^2 \right)^{5b 3 \nu} A^b_\nu = 
	- \left( m^2 \right)^{7b 2 \nu} A^b_\nu = 
	- \left( m^2 \right)^{7b 3 \nu} A^b_\nu = 
\nonumber \\
	&&
	3 g \mathcal B^2_0 \left( 
		2 \Delta_{11} + 2 \Delta_{33} + \Delta_{44} 
	\right) \frac{-1 + f}{r^2} \phi^2 = \frac{2 m^2}{g} 
	\frac{1 - f}{r^2} \phi^2 , 
\label{2-50}\\
	\left( \mu^2 \right)^{4b 3 \nu} A^b_\nu &=& 
		- \left( \mu^2 \right)^{5b 2 \nu} A^b_\nu = 
		\left( \mu^2 \right)^{5b 3 \nu} A^b_\nu = 
		- \left( \mu^2 \right)^{7b 2 \nu} A^b_\nu = 
		- \left( \mu^2 \right)^{7b 3 \nu} A^b_\nu = 
\nonumber \\
	&&
	\frac{g}{2} \left( \mathcal A^2_0 - \mathcal A^2_2 \right) 
	\left( \delta_5 + \delta_7 \right) 
	\frac{-1 + f}{r^2} = \frac{2 \mu^2}{g} 
	\frac{1 - f}{r^2} , 
\label{2-60}\\
	\frac{\left( m^2_\phi \right)^{ab \mu \nu} A^a_\nu A^b_\mu}
		{G^{mm \alpha}_{\phantom{mm} \alpha}} &=& 
	4 \frac{\left( 1 - f \right)^2 }{r^2} . 
\label{2-70}
\end{eqnarray}
Finally we have following set of equations for the unknown functions $f(r)$ and $\phi(r)$ 
\begin{eqnarray}
	- f^{\prime \prime} + \frac{f \left( f^2 - 1 \right) }{x^2} - 
	m^2 \left( 1 - f \right) \phi^2 &=& - \mu^2 \left( 1 - f \right) , 
\label{2-80}\\
	\phi^{\prime \prime} + \frac{2}{x} \phi^\prime &=& 
	\phi \left[ 
		m^2 \frac{\left( 1 - f \right)^2}{x^2} + 
		\lambda \left( 
			\phi^2 - M^2
		\right) 
	\right] 
\label{2-90}
\end{eqnarray}
where we have introduced dimensionless variable $x = r/r_0$ and redefine 
$r_0 \phi \rightarrow \phi$, $r_0 M \rightarrow M$; $r_0 = \phi(0)^{-1}$. 

\begin{figure}[h]
	\begin{minipage}[ht]{.45\linewidth}
		\begin{center}
			\fbox{
				\includegraphics[width=.9\linewidth]{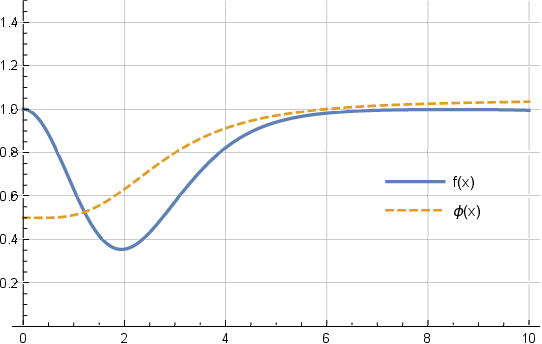}
			}
		\end{center}
		\caption{The profiles of eigenfunctions $f(x)$ and $\phi(x)$. Eigenvalues 
		$\mu^2 = 2.609393$, $M^2 = 1.03243$, $m = 2$, $\lambda = 0.1$, 
		$\phi(0) = 0.5$, $f_2 = -1.0$. 
			}
	\label{functions}
	\end{minipage}\hfill
	\begin{minipage}[ht]{.45\linewidth}
		\begin{center}
			\fbox{
				\includegraphics[width=.9\linewidth]{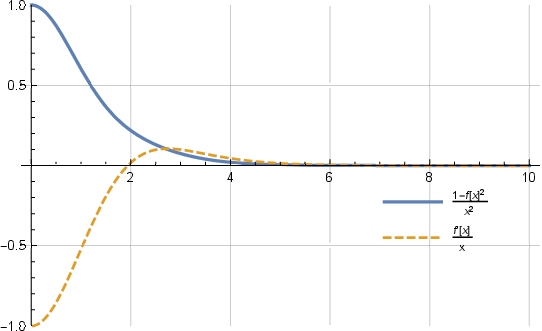}
			}
		\end{center}
		\caption{The profiles of magnetic fields $H^{2,5,7}_r(x)$ and 
		$H^{2,5,7}_{\theta, \varphi}(x)$.
		}
		\label{fields}
	\end{minipage}
\end{figure}

The numerical solution of equations set \eqref{2-80} \eqref{2-90} is considered as an nonlinear eigenvalue problem with eigenfunctions $f, \phi$ and eigenvalues $\mu, M$. The boundary conditions are 
\begin{equation}
	f(0) = 1, f^\prime(0) = 0, \phi(0) = 0.5, \phi^\prime(0) = 0 . 
\label{2-100}
\end{equation}
The results of numerical calculations are presented in Fig. \ref{functions}. 
One can find the approximate solution close to origin:
\begin{eqnarray}
	f(x) &=& 1 + f_2 \frac{x^2}{2} + \ldots , 
\label{2-110}\\
	\phi (x) &=& \phi_0 + \phi_2 \frac{x^2}{2} + \ldots 
\label{2-120}
\end{eqnarray}
where $f_2$ is arbitrary and $\phi_2 = \lambda \phi_0 (\phi_0^2 - M^2)/3$. The asymptotic behavior can be obtained directly from equations \eqref{2-80} \eqref{2-90} and are 
\begin{eqnarray}
	f(x) &\approx& 1 - f_\infty e^{-x \sqrt{m^2 M^2_\infty - \mu^2}} , 
\label{2-130}\\
	\phi (x) &\approx& M - 
 	\phi_\infty \frac{e^{- x \sqrt{2 \lambda M^2}}}{x} 
\label{2-140}
\end{eqnarray}
here $f_\infty, \phi_\infty$ are constants. 
The field strength $F^a_{ij}, i,j = r, \theta, \phi$ is 
\begin{eqnarray}
	F^2_{ij} &=& \frac{1}{g}
	\begin{pmatrix}
		0	& 0 & 2 \sin^2 \theta f^\prime \\
		0	& 0 & \sin 2 \theta \left( -1 + f^2 \right) \\
		- 2 \sin^2 \theta f^\prime	& - \sin 2 \theta \left( -1 + f^2 \right) & 0 
	\end{pmatrix} , 
\label{2-150}\\
	F^5_{ij} &=& \frac{1}{g}
	\begin{pmatrix}
		0	& - 2\cos \varphi f^\prime & \sin 2 \theta \sin \varphi f^\prime \\
		2\cos \varphi f^\prime	& 0 & - 2 \sin^2 \theta \sin \varphi \left( -1 + f^2 \right) \\
		- \sin 2 \theta \sin \varphi f^\prime	& 2 \sin^2 \theta \sin \varphi \left( -1 + f^2 \right) & 0 
	\end{pmatrix} , 
\label{2-160}\\
	F^7_{ij} &=& \frac{1}{g}
	\begin{pmatrix}
		0	& - 2 \sin \varphi f^\prime & - \sin 2 \theta \cos \varphi f^\prime \\
		2 \sin \varphi f^\prime	& 0 & 2 \sin^2 \theta \cos \varphi \left( -1 + f^2 \right) \\
		\sin 2 \theta \cos \varphi f^\prime	& - 2 \sin^2 \theta \cos \varphi \left( -1 + f^2 \right) & 0 
	\end{pmatrix} .
\label{2-170}
\end{eqnarray}
In other words we have the following chromomagnetic fields: 
\begin{eqnarray}
	H^{2,5,7}_{\theta, \varphi} &\propto& \frac{f^\prime}{r} , 
\label{2-180}\\
	H^{2,5,7}_r &\propto& \frac{1 - f^2}{r^2}. 
\label{2-190}
\end{eqnarray}
Taking into account the asymptotic behavior \eqref{2-130} we have the following asymptotic behavior of magnetic fields 
\begin{eqnarray}
	H^{2,5,7}_{\theta, \varphi}(r) &\approx& \frac{1}{g} 
	\frac{e^{-\frac{r}{r_0} \sqrt{m^2 M^2 - \mu^2}}}{r r_0} , 
\label{2-200}\\
	H^{2,5,7}_r(r) &\approx& \frac{1}{g} 
	\frac{e^{-\frac{r}{r_0} \sqrt{m^2 M^2 - \mu^2}}}{r^2} . 
\label{2-210}
\end{eqnarray}
This asymptotic behavior shows us that we have a monopole but with exponential decreasing radial magnetic field $H^a_r$. The reason for this exponential decrease is the quantum corrections terms, $M, \mu$ in equations \eqref{2-80} \eqref{2-90}.

The reason of such decreasing is quantum corrections $m, M, \mu$. 

The problem of justification of our two equations approximation applied for quantum monopole is very hard and for solving this problem we have to compare  our results with lattice calculations. The asymptotic behavior \eqref{2-210} allows us to calculate the magnetic flux $\Phi(r)$ through a sphere of radius $r$ around quantum monopole:
\begin{equation}
	\Phi(r) = 4 \pi r^2 H_r \approx 4 \pi 
	e^{-\frac{r}{r_0} \sqrt{m^2 M^2 - \mu^2}} 
\label{2-220}
\end{equation}
that coincides with the lattice calculations of the magnetic flux in \cite{Bornyakov:2003vx}. Equation \eqref{2-220} allows us to estimate the magnetic screening length of the chromomagnetic field $H_r$ as 
\begin{equation}
	\xi = \frac{\sqrt{m^2 M^2 - \mu^2}}{r_0} .
\label{2-230}
\end{equation}
The physical meaning of the obtained solution is as follows. In this situation we have the Meissner effect: the condensate $\phi$ pushes out the chromomagnetic fields $H^a_i$ into a spherical bag creating a quantum monopole. It is widely believed that the pair monopole - antimonopole will form an object similar to Cooper pair in superconductivity. In order to have such object a flux tube filled with the chromomagnetic field should be stretched between monopole and antimonopole. In the next section we want to show that such a flux tube actually arises in the condensate $\phi$ that is also the manifestation of the Meissner effect in QCD. 

\section{Flux tube filled with chromomagnetic field}
\label{sec4}

In this section we will obtain a flux tube filled with chromomagnetic field and created by two infinitely separated monopole and antimonopole. We consider the  infinite flux tube because the consideration of a finite flux tube is a much more difficult problem since we have to consider a non-linear eigenvalue problem with partial differential equations. 

We use the same equations \eqref{1-80} \eqref{1-90} and the flux tube ans\"atz of the form 
\begin{eqnarray}
	A^2_\varphi &=& \frac{\rho}{g} w(\rho),
\label{3-10} \\ 
	\phi &=& \phi(\rho)
\label{3-20}
\end{eqnarray} 
The matrix $\Delta^{AB}$ is chosen in the form \eqref{2-20} with arbitrary $\Delta_{11}, \Delta_{33}, \Delta_{44}, \Delta_{66}, \Delta_{88}$ and vectors $\mathcal A_\mu, \mathcal B_\mu$ 
\begin{eqnarray}
	\mathcal A_\mu &=& \left( 
		\mathcal A_0, \mathcal A_1, \mathcal A_2, 0 
	\right) , 
\label{3-30}\\
	\mathcal B_\mu &=& \left( 
		\mathcal B_0, \mathcal B_1, 0, 0 
	\right)  .
\label{3-40}
\end{eqnarray}
Using these $\Delta^{AB}$, $\mathcal A_\mu$ and $\mathcal B_\mu$ we obtain 
\begin{eqnarray}
	\left( m^2 \right)^{2b \varphi \nu} A^b_\nu &=& 
	\frac{3g}{4} \left( 
		\mathcal B_0^2 - \mathcal B_1^2 
	\right) 
	\left( 
		4 \Delta_{11} + 4 \Delta_{33} + \Delta_{44} + \Delta_{66} 
	\right) \frac{w}{\rho} \phi^2 = 
	- \frac{m^2}{g} \frac{w}{\rho} \phi^2 , 
\label{3-50}\\
	\left( \mu^2 \right)^{2b \varphi \nu} A^b_\nu &=& 
	\frac{g}{4} \left( 
			\mathcal A_0^2 - \mathcal A_1^2 - \mathcal A_2^2 
		\right) 
		\left( 
				\delta_{5} + \delta_{7} 
			\right) \frac{w}{\rho} = 
		- \frac{\mu^2}{g} \frac{w}{\rho} , 
\label{3-60}\\
	\frac{\left( m^2_\phi \right)^{ab \mu \nu} A^a_\nu A^b_\mu}
	{G^{mm \alpha}_{\phantom{mm} \alpha}} &=& - 
	\frac{4 \Delta_{11} + 4 \Delta_{33} + \Delta_{44} + \Delta_{66}}
	{4 \left( 
		\Delta_{11} + \Delta_{33} + \Delta_{44} + \Delta_{66} + \Delta_{88}
	\right)} w^2 . 
\label{3-70}
\end{eqnarray}
That leads to the set of equations  
\begin{eqnarray}
	- w'' - \frac{w'}{x} + \frac{w}{x^2} + m^2 \phi^2 w &=& 
	\mu^2 w , 
\label{3-80}\\
	\phi^{\prime \prime } + \frac{\phi'}{x} &=& 
	\phi \left[ 
		m^2_\phi w^2 + \lambda \left( 
			\phi^2 - M^2
		\right) 
	\right] . 
\label{3-90}
\end{eqnarray}
here we have introduced the dimensionless variable $x = \rho/\rho_0$ and redefine 
$\rho_0^2 w \rightarrow w$, $\rho_0 \phi \rightarrow \phi$, 
$\rho_0 M \rightarrow M$, $m_\phi / \rho_0 \rightarrow m_\phi$; 
$\rho_0 = \phi(0)^{-1}$. 

\begin{figure}[h]
	\begin{minipage}[ht]{.45\linewidth}
		\begin{center}
			\fbox{
				\includegraphics[width=.9\linewidth]{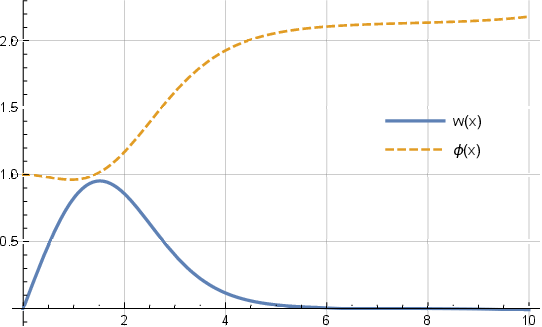}
			}
		\end{center}
		\caption{The profiles of eigenfunctions $w(x)$ and $\phi(x)$. Eigenvalues 
		$\mu = 1.562199$, $M = 2.131698$, $m = m_\phi = w_1 = 1$, $\phi(0) = 1$. 
			}
	\label{FTfunctions}
	\end{minipage}\hfill
	\begin{minipage}[ht]{.45\linewidth}
		\begin{center}
			\fbox{
				\includegraphics[width=.9\linewidth]{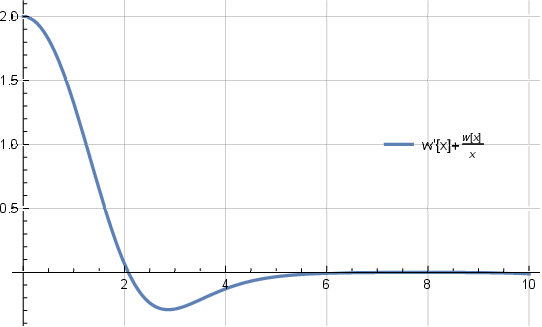}
			}
		\end{center}
		\caption{The profile of longitudinal chromomagnetic field $H^2_z$. 
		}
	\label{FTfields}
	\end{minipage}
\end{figure}

Again we solve set of equations as a non-linear eigenvalue problem with 
$w, \phi$ as eigenfunctions and $M, \mu$ eigenvalues. The boundary conditions are 
\begin{equation}
	w(0) = 0, w'(0) = w_1, \phi(0) = 1, \phi'(0) = 0. 
\label{3-100}
\end{equation}
The results of numerical calculations are presented in Fig. \ref{FTfunctions}. One can find the approximate solution close to origin:
\begin{eqnarray}
	w(x) &=& w_1 x + w_3 \frac{x^3}{3!} + \ldots , 
\label{3-102}\\
	\phi (x) &=& \phi_0 + \phi_2 \frac{x^2}{2} + \ldots 
\label{3-104}
\end{eqnarray}
where $w_1$ is arbitrary and $\phi_2 = \lambda \phi_0 (\phi_0^2 - M^2)/2$. The asymptotic behavior can be obtained directly from equations \eqref{3-80} \eqref{3-90} and are 
\begin{eqnarray}
	w(x) &\approx& w_\infty 
	\frac{e^{-x \sqrt{m^2 M^2 - \mu^2}}}{\sqrt{x}} , 
\label{3-106}\\
	\phi (x) &\approx& M - 
 	\phi_\infty \frac{e^{- x \sqrt{2 \lambda M^2}}}{\sqrt{x}} 
\label{3-108}
\end{eqnarray}
here $w_\infty, \phi_\infty$ are constants. 

Chromomagnetic field for the potential \eqref{3-10} is 
\begin{equation}
	H^2_z = \frac{1}{g}\left( 
		w' + \frac{w}{\rho}
	\right) 
\label{3-110}
\end{equation}
and is presented in Fig. \ref{FTfields}. 

\section{Meissner effect}
\label{sec5}

In this section we want to discuss the physical meaning of the obtained solutions for the quantum monopole and flux tube. The question is: Are the quantum monopole and flux tube embedded in empty space or into a condensate with non-zero energy density ? In order to understand this we have to analyze the asymptotic behavior of the corresponding energy densities for the quantum monopole and flux tube. To solve this question we will write the Lagrangian for the set of equations \eqref{2-80} \eqref{2-90} (monopole equations) and \eqref{3-80} \eqref{3-90} (flux tube equations). Since we consider the static case then the Lagrangian density coincides with the energy density. We will require that the energy density should be everywhere positive. 

\subsection{Monopole case}

For the monopole equations \eqref{2-80} \eqref{2-90} we can write following Lagrangian 
\begin{equation}
	L_{mnpl} = \epsilon_{mnpl} = \frac{{f'}^2}{2 r^2} + \frac{{\phi'}^2}{2} + 
	\frac{\left( f^2 - 1 \right)^2}{4 r^4} + 
	\frac{m^2}{2} \frac{\left( f - 1 \right)^2}{r^2} \phi^2 - 
	\frac{\mu^2}{2 r^2} \left( f - 1 \right)^2 + 
	\frac{\lambda}{4} \left( 
		\phi^2 - M^2
	\right)^2 + \epsilon_\infty . 
\label{5-1-10}
\end{equation}
Since we consider the static case then \eqref{5-1-10} is the energy density also. In order to have positive energy density in the whole space we have to add some constant $\epsilon_\infty$. The profile of the dimensionless energy density $\epsilon_{mnpl}(x)$ for the solution presented in Fig. \ref{functions} is presented in Fig. \ref{energy_densities}. 

\subsection{Flux tube case}

\begin{figure}[h]
	\begin{center}
		\fbox{
			\includegraphics[width=.5\linewidth]{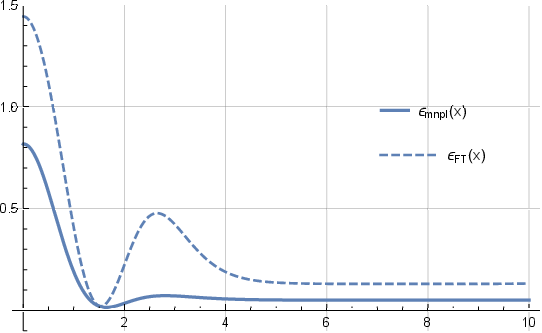}
		}
	\end{center}
	\caption{The profiles of energy densities for quantum monopole and flux tube. $\epsilon_\infty \approx 0.05$ for the monopole and 
	$\epsilon_\infty \approx 0.13$ for the flux tube. 
	}
\label{energy_densities}
\end{figure}

For the flux tube equations \eqref{3-80} \eqref{3-90} we have the following Lagrangian 
\begin{equation}
	L_{ft} = \epsilon_{ft} = \frac{{w'}^2}{2} + \frac{{\phi'}^2}{2} + 
	\frac{w^2}{2 r^2} + \frac{m^2}{2} w^2 \phi^2 - 
	\frac{\mu^2}{2} w^2 + 
	\frac{\lambda}{4} \left( 
		\phi^2 - M^2
	\right)^2 + \epsilon_\infty . 
\label{6-1-10}
\end{equation}
In order to obtain set of equations \eqref{3-80} \eqref{3-90} we consider the case $m_\phi = m$. Once again \eqref{5-1-10} is the energy density also, and we add some constant $\epsilon_\infty$ to have positive energy density in the whole space. The profile of the dimensionless energy density $\epsilon_{ft}(x)$ for the solution presented in Fig. \ref{FTfunctions} is presented in Fig.~\ref{energy_densities}. 

Analyzing the profiles of both energy densities at the infinity we can say that our monopole and flux tube are embedded into a quantum condensate formed by coset fields $A^m_\mu$. 

\section{Discussion and conclusions}

In this letter we have shown that in two equations approximation we can obtain quantum monopole and flux tube both embedded into a quantum condensate formed by quantum gauge fields $A^m_\mu$. Both objects are constructed from gauge fields $A^a_\mu \in SU(2) \subset SU(3)$. The fields $A^a_\mu$ are pushed out by coset fields $A^m_\mu$ that is the manifestation of the Meissner effect in QCD. The subgroup fields $A^a_\mu$ have non-zero expectation values and quantum oscillation around these mean values. The coset fields $A^m_\mu$ are pure quantum ones with zero expectation values. The equations describing both objects contain quantum corrections arising from the dispersion of quantum fluctuations $A^{a,m}_\mu$ fields. 

The quantum monopole contains radial chromomagnetic field which exponentially decreases at infinity due to the presence of the quantum corrections coming from the dispersion of quantum fluctuations $A^a_\mu$ fields. We have shown that in this situation we have the Meissner effect: the magnetic fields are pushed out by the condensate $\phi$. 

It is widely believed that such monopole - anti-monopole pair is connected by flux tube filled with the longitudinal chromomagnetic field. We confirmed that in two equations approximation such infinite flux tube really exists. We have shown that the tube is filled by chromomagnetic field which is pushed out from the condensate $\phi$ due to the Meissner effect in QCD. 

The problem of justification of our two equations approximation applied for obtaining quantum monopole and flux tube is very hard and for solving this problem we have to compare our results with lattice calculations. And even more: such comparing is very interesting and can be considered in future in an independent paper. 

Finally, we have shown the existence of :
\begin{itemize}
\item quantum monopole in two equations approximation;
\item flux tube with longitudinal chromomagnetic field in two equations approximation;
\item Meissner effect in two equations approximation. 
\end{itemize}

\section*{Acknowledgements}

I acknowledge a grant 
in fundamental research in natural sciences by the Ministry of Education and Science of Kazakhstan. I am very grateful for D. Singleton for fruitful discussion, comments and criticism. 



\end{document}